\documentclass[prd,preprint,a4paper,superscriptaddress,nofootinbib,11pt]{revtex4}

\usepackage{amsfonts}
\usepackage{amssymb}
\usepackage{hyperref}

\def\fs{\footnotesize}
\def\uz{\ ^{(0)}}
\def\uu{\ ^{(1)}}

\begin{document}


\title{\Large Post-Newtonian Parameters from Alternative Theories of Gravity}
\date{\today}

\author{Gianluca ALLEMANDI}
\email{allemandi@dm.unito.it}
\affiliation{\fs Dipartimento di Matematica,  Universit\`a di Torino, Via C. Alberto 10, 10123
Torino and INFN Torino}
\author{Mauro FRANCAVIGLIA}
\email{francaviglia@dm.unito.it}
\affiliation{\fs Dipartimento di Matematica,  Universit\`a di Torino, Via C. Alberto 10, 10123
Torino and INFN Torino}
\author{Matteo Luca RUGGIERO}
\email{matteo.ruggiero@polito.it}
\affiliation{\fs Dipartimento di Fisica, Politecnico di Torino, Corso Duca degli Abruzzi 24, 10129 Torino and INFN Torino}
\author{Angelo TARTAGLIA}
\email{angelo.tartaglia@polito.it}
\affiliation{\fs Dipartimento di Fisica, Politecnico di Torino, Corso Duca degli Abruzzi 24, 10129 Torino and INFN Torino}


\begin{abstract}
Alternative theories of gravity have been recently studied in
connection with their cosmological applications, both in the
Palatini and in the metric formalism. The aim of this paper  is to
propose a theoretical framework (in the Palatini formalism) to
test these theories at the solar system level and possibly at the
galactic scales. We exactly solve field equations in vacuum and
find the corresponding corrections to the standard general
relativistic gravitational field. On the other hand, approximate
solutions are found in matter cases starting from a Lagrangian
which depends on a phenomenological parameter. Both in the vacuum
case and in the matter case the deviations from General Relativity
are controlled by parameters that provide the Post-Newtonian
corrections which prove to be  in good agreement with solar system
experiments.

\end{abstract}

\maketitle

\section{Introduction}\label{sec:intro}
The most striking and recent experimental discovery regarding
Cosmology  and the structure of the universe is related with the
evidence of the acceleration of the universe, which is supported
by experimental data deriving from different tests: i.e., from
Type-Ia Supernovae, from CMWB and from the large scale structure of the universe \cite{Perlmu}. Standard General Relativity is not able to provide a theoretical explanation to these experimental results unless some {\it exotic and invisible matter} is admitted to exist in the universe ({\it Dark Energy}). Proposals to explain the cosmic acceleration also  arise from higher dimensional theories of Gravity \cite{dvali}.  Alternative theories to explain the acceleration of the universe have been recently proposed in the framework of higher order theories of Gravity \cite{carrol2}, already introduced in the framework of cosmological models to explain the early time inflation \cite{staro}.
Different models have been  then studied both in the standard metric formalism \cite{metricfR} and in the first order Palatini formalism  \cite{palatinifR}. Higher order theories of Gravity have been studied also in a quantum framework and a quantization of $L(R)$ theories has been performed in \cite{zerbo}. \\
To test the theoretical consistence of these theories with
observational data is however necessary to examine and to fit the
standard tests for General Relativity: in particular solar system
experiments and the tests of gravity at galactic scales.
General Relativity reproduces with an excellent precision the experimental results obtained at
the solar system scale \cite{solar}. This naturally implies that each  theory which  pretends
to be consistent with experimental results should surely reproduce General Relativity in this limit.\\
The aim of this paper is to provide a general theoretical
framework to test the reliability of alternative theories of
Gravity with solar system experiments.   Such a problem  was
already studied from a different viewpoint in the standard metric
formalism in \cite{solarmet} and in the Palatini formalism in
\cite{palatinifR}, \cite{solarpal}. Some debate is still open on
the accordance of experimental results with solar system
experiments and some authors erroneously  claim that only theories
which do not differ too much from General Relativity do the job
(see \cite{palatinifR}); however, as we shall see, this is not
true and, moreover, it is known that the Palatini formalism can
naturally provide accordance with solar system experiments (see
e.g. \cite{solarpal}, \cite{Barraco}, \cite{Barraco2} and
references quoted therein). Some interesting results are also
present in literature regarding the accordance of alternative
theories of Gravity with rotational curves of galaxies
\cite{capogal}. In this paper we shall study the problem of the
reliability of alternative theories of Gravity with solar system
experiments and give also some hints regarding the galactic scale
tests of Gravity (which will be considered in a forthcoming paper
\cite{MGnew}). We shall do this  from a purely theoretical
viewpoint, trying to understand which Newtonian or Post-Newtonian modifications to standard General Relativity arise from specific modifications of the Hilbert-Einstein Lagrangian. In particular we consider $L(R)$ theories where the Lagrangian depends on an arbitrary analytic function $L$ of the  scalar curvature $R$. Starting from the results already obtained in \cite{buchdahl} and \cite{FFV}  we find an exact solution to field equations in vacuum. In that case field equations are controlled by a scalar-valued equation called the {\it structural equation}. It is relevant  that modifications to the standard general relativistic gravitational field arise,  and they turn out to be directly related to solutions of the structural equation and, consequently, to the particular form of the Lagrangian chosen (the choice of $L(R)$). We shall show how these modifications can be suitably  interpreted as Post-newtonian parameters related to the non linearity of the theory. \\
We consider furthermore field equations in the case of matter universes (i.e. when the stress energy tensor is non vanishing). Considering a linear approximation of the metric, either with respect to a Minkowski flat space-time, or with respect to a de Sitter or an anti de Sitter space-times, the non-linear structure of the theory influences the gravitational field. We stress however that, in the first order approximation of the Palatini formalism, the presence of non-linear terms in the Lagrangian only influences the definition of $R$, while field equations remain unchanged. We finally derive the gravitational field for the particular Lagrangians $R+ \alpha f(R)$ where $\alpha$ is an adimentional parameter. The corresponding gravitational potential contains then a term which is directly proportional to $\alpha$, such that General Relativity is reproduced in the limit $\alpha=0$, as it should be expected. This implies that the parameter $\alpha$ behaves as a sort of scale parameter which becomes relevant at large scales and it can be interpreted as a Post-Newtonian parameter ensuing from the non linearity of the Lagrangian.\\
Our approach,  of course does not  completely solve the problem of the generic reliability of alternative theories of gravity at solar system and galactic scale. However by introducing some Post-Newtonian parameters, it shows that General Relativity is certainly reproduced at small scales (as it is expected) for large families of Lagrangians.
Further comparisons with other classical tests of General Relativity  and applications to more general cases, as well as tests of Gravity at large (galactic) scale will be presented in the forthcoming paper \cite{MGnew}.


\section{The theoretical  framework of $L(R)$ gravity}\label{sec:frgravity}

\noindent We deal with a $4$-dimensional gravitational theory on a Lorentzian manifold ($M,g$) with signature $(-,+,+,+)$\footnote{If not otherwise stated, we use units such that $G=c=1$.}.
The action is chosen to be:
\begin{equation}
A=A_{\mathrm{grav}}+A_{\mathrm{mat}}=\int [ \sqrt{g} L (R)+2\kappa L_{\mathrm{mat}} (\psi, \nabla \psi) ]  \; d^{4}x
\end{equation}
where $R\equiv R( g,\Gamma) =g^{\alpha\beta}R_{\alpha \beta}(\Gamma )$, $
R_{\mu \nu }(\Gamma )$ being the Ricci tensor of any torsionless connection $\Gamma$   independent on a metric  $g$
 is assumed to be the physical metric.
The gravitational part of the Lagrangian is represented by any real analytic function $L (R)$ of one real variable, which is assumed
to be the scalar curvature $R$.
The total Lagrangian contains also a first order matter part
$L_{\mathrm{mat}}$ functionally depending on yet unspecified  matter fields $\Psi$ together with their first derivatives,
equipped with a gravitational coupling constant $\kappa=\frac{8\pi G}{c^4}$ (see e.g. \cite{buchdahl}).\\
Equations of motion ensuing from the first order \'a la Palatini formalism
are  (see \cite{Barraco, palatinifR, FFV})
\begin{eqnarray}
L^{\prime }(R) R_{(\mu\nu)}(\Gamma)-\frac{1}{2} L(R)  g_{\mu \nu
}&=&\kappa T_{\mu \nu }^{mat}  \label{ffv1}\\
\nabla _{\alpha }^{\Gamma }[ \sqrt{g} L^\prime (R) g^{\mu \nu })&=&0
\label{ffv2}
\end{eqnarray}
where $T^{\mu\nu}_{mat}=-\frac{2}{\sqrt g}\frac{\delta L_{\mathrm{mat}}}{\delta g_{\mu\nu}}$
denotes the matter source stress-energy tensor and $\nabla^{\Gamma}$ means covariant derivative with respect to the connection $\Gamma$, which we recall to be independent on the metric $g$. In this paper  the metric $g$ and its inverse
are used for lowering and raising indices.\\
We denote by $R_{(\mu\nu)}$ the symmetric part of $R_{\mu\nu}$, i.e. we set
$R_{(\mu\nu)}\equiv \frac{1}{2}(R_{\mu\nu}+R_{\nu\mu})$. From (\ref{ffv2}) it follows that $\sqrt{g} L^\prime (R)  g^{\mu \nu }$
is a symmetric twice contravariant tensor density of weight $1$, so that it can be used (if non degenerate) to define a new metric $h_{\mu \nu}$ by the prescription:
\begin{equation}\label{h_met}
\sqrt{g} L^{\prime } (R)  g^{\mu \nu}=\sqrt{h} h^{\mu \nu}
\end{equation}
which is generically invertible. This means that the two metrics $h$ and $g$ are conformally equivalent so that
space-time $M$ can be a posteriori endowed with a bi-metric structure $(M,g,h)$ \cite{FFV} equivalent to the original metric-affine structure $(M,g,\Gamma)$. The
corresponding conformal factor can be easily found to be $L^{\prime} (R) $, since (\ref{h_met}) gives:
\begin{equation}\label{h_met2}
h_{\mu \nu }=L^\prime (R)  \;  g_{\mu \nu }
\end{equation}
Therefore, as it is well known, equation (\ref{ffv2}) implies that  $\Gamma =\Gamma _{LC}(h)$, i.e. the dynamical connection turns out a posteriori to be the Levi-Civita connection of the newly defined metric $h$, so that $R_{(\mu\nu)}(\Gamma_{LC} (h) )=R_{\mu \nu }(h)\equiv R_{\mu\nu}$ is now the metric Ricci tensor of the new metric $h$. \\
Equation (\ref{ffv1}) can be supplemented by the scalar-valued equation
obtained by taking the $g$-trace of (\ref{ffv1}), where we set $\tau=\mathrm{tr} T=g^{\mu \nu }T^{mat}_{\mu \nu }$:
\begin{equation} \label{struct}
L^{\prime} (R) R-2 L(R)= \kappa\tau
\end{equation}
Equation (\ref{struct}) is called the \textit{structural equation} and it controls the solutions of equation (\ref{ffv1}). For any real solution $R=F(\tau)$ of (\ref{struct}) we have  in fact that  both $L(R)=L(F(\tau))$ and $L^\prime(R)=L^\prime(F(\tau))$ can be seen as functions of $\tau$. For notational convenience we shall use the abuse of notation $L(\tau)=L(F(\tau))$ and $L^\prime(\tau)=L^\prime(F(\tau))$.\\
Now we are in position to introduce the generalized Einstein
equations under the  form\footnote{Provided that $L^\prime(\tau)
\neq 0$: see below.}
\begin{equation}\label{gen_Ein}
R_{\mu \nu }\left( h\right)=\frac{L(\tau)}{2  L^{\prime} (\tau) }
 g_{\mu \nu }+\frac{\kappa }{L^{\prime} (\tau) }T_{\mu \nu }
\end{equation}
with $h_{\mu\nu}$ defined by (\ref{h_met2}) for a given $g_{\mu \nu }$ and $T^{mat}_{\mu\nu}$
(see also \cite{Barraco, FFV, palatinifR}).

\section{Some exact Solution of the Field Equations in vacuum} \label{sec:exactvacuum}

In this Section we look for a spherically symmetrical solution of
the generalized Einstein equations in vacuum, starting from the
results obtained in \cite{buchdahl} and \cite{FFV}. To this end first notice that eqs. (\ref{ffv1}-\ref{ffv2}), in vacuum,
can be written under the form
\begin{eqnarray}
[L'(R)] R_{(\mu\nu)}(\Gamma)-\frac{1}{2}[L(R)] g_{\mu \nu
}&=&0  \label{ffv111}\\
\nabla _{\alpha }^{\Gamma }(\sqrt{g} \; [L'(R)] \; g^{\mu \nu
})&=&0 \label{ffv211}
\end{eqnarray}
Furthermore, the structural equation (\ref{struct})
becomes
\begin{equation}
 L^{\prime }(R) R-2L(R)= 0 \label{eq:fvac2}
\end{equation}
In order to solve (\ref{ffv111})-(\ref{ffv211}), we follow the
discussion outlined in \cite{FFV}. Let us suppose that the
structural equation (\ref{eq:fvac2}) is not identically satisfied
and has a countable set of (real) solutions ($i=1,2....$):
\begin{equation}
R=c_i \label{eq:solerre}
\end{equation}
Then, we have two possibilities, depending on the value of the
first derivative $L'(R)$ evaluated at the point $R=c_i$:
\begin{enumerate}
\item $L'(c_i) = 0$ \\
\item $L'(c_i) \neq 0$
\end{enumerate}
In the first case, eqs. (\ref{eq:fvac2}) implies that also
$L(c_i)=0$, and, hence, the equations of motion
(\ref{ffv111})-(\ref{ffv211}) are identically satisfied.  The only
relation between $g$ and $\Gamma$ is the following
\begin{equation}
R(g,\Gamma)=c_i \label{eq:relggamma}
\end{equation}
Indeed, this equation is not sufficient in this case to determine an explicit
relation between the metric and the connection. Hence, in what
follows, we shall suppose that $L'(c_i) \neq 0$. \\
We remark that if the Lagrangian is in the form $L(R)=R^n$, with $n \geq 2$, $n \in \mathbb{N}$, $R=0$ is solution of
eq. (\ref{eq:fvac2}), and, moreover one has $L'(R=0)=0$. Consequently we exclude such Lagrangians.\\
If $L'(c_i) \neq 0$ then the solution of the equations of motion
(\ref{ffv111})-(\ref{ffv211}) is given by the Levi-Civita
connection of the metric $h$, which, in turn turns out to be
equivalent to Levi-Civita connection of the physical metric $g$
(owing to the relation $h=L'(c_i)g$). Accordingly, the metric $g$ is  the solution of the generalized
Einstein equations
\begin{equation}
R_{\mu \nu }\left( g\right)=\mu
 g_{\mu \nu }
  \label{eq:fvac4}
\end{equation}
where
\begin{equation}
\mu=c_i/4 \label{eq:defk1}
\end{equation}

We look for a static solution of the field equations
(\ref{eq:fvac4}) describing the field outside a spherically
symmetric mass distribution. Hence we may write the metric in the
form
\begin{equation}
ds^2=-e^{\Phi(r)}dt^2+e^{\Lambda(r)}dr^2+r^2d\vartheta^2+r^2\sin^2
\vartheta d\varphi^2 \label{eq:metrica1}
\end{equation}
It is easy to check that the field equations (\ref{eq:fvac4}) are
satisfied if we set
\begin{equation}
-e^{\Phi(r)}=g_{tt}= -1 +\frac{C}{r}-\frac{\mu r^2}{3} \label{eq:gtt}
\end{equation}
and
\begin{equation}
e^{\Lambda(r)}=g_{rr}= \left(1
-\frac{C}{r}+\frac{\mu r^2}{3}\right)^{-1} \label{eq:grr}
\end{equation}
where $C$ is an arbitrary constant; in particular, the metric
defined by (\ref{eq:gtt})-(\ref{eq:grr}) corresponds to the so
called \textit{Schwarzschild-de Sitter space-time} (see
\cite{he},\cite{ruffi}). The physical meaning of the constant $C$
becomes clear when considering the limit of weak gravitational
field. We know that in General Relativity in this limit we have
\begin{equation}
g_{tt} \simeq -\left(1+2\phi \right) \label{eq:newtlimit1}
\end{equation}
where
\begin{equation}
\phi=-\frac{M}{r} \label{eq:newtlimit2}
\end{equation}
is the Newtonian  potential,
$M$ being the mass of the spherically symmetric source of the
gravitational field. Consequently, in order to obtain the Newtonian limit we must set $C=2M$. Moreover, from (\ref{eq:gtt}) it
is evident that  a further contribution to the standard Newtonian
potential is present in higher order theories of gravity. In
particular, this contribution is proportional to the values of the
Ricci scalar, owing to the proportionality between $\mu$ and $c_i$
(see (\ref{eq:solerre}) and (\ref{eq:defk1})). This implies that the
higher order contribution to the gravitational potential should be
small enough not to contradict the known tests of gravity. In the
case of small values of $R$ (which surely occur at solar system
scale) the Einsteinian limit (i.e. the Schwarzschild solution) and
 the Newtonian limit are recovered, as it is evident from
(\ref{eq:gtt}). In this context, $\mu$ can be naturally thought of as a Post-Newtonian parameter,
 ensuing from the non linearity of the theory ($\mu=0$ for the Hilbert-Einstein Lagrangian).\\
On the other hand this Post-Newtonian correction could play  some role at larger scales and it
could be interesting to test higher-order theories at galactic scales, as already done in the metric formalism in \cite{capogal}.


\section{Field Equations in linear approximation}\label{sec:felinear}

We aim at writing the field equations for  Lagrangians $L(R)=R+\alpha f(R)$  in linear
approximation: that is, we are going to solve the field equation
at first order approximation with respect to a given background.
In other words, we suppose to know a background solution of field equations
(\ref{ffv1}), (\ref{ffv2}) determined by the
affine connection $^{(0)}\Gamma$ and the metric $^{(0)}g$ \footnote{Here and henceforth, the superscripts $^{(0)}$ and $^{(1)}$ refer to the background and perturbed quantities, respectively.}. We now
perturb this solution by writing
\begin{eqnarray}
\Gamma_{\ \mu\nu}^{\alpha} & = & ^{(0)}\Gamma_{\
\mu\nu}^{\alpha}+^{(1)}\Gamma_{\ \mu\nu}^{\alpha}
\label{eq:gammadef} \\
g_{\mu\nu} & = &  ^{(0)}g_{\mu\nu}+^{(1)}g_{\mu\nu}
\label{eq:gmunudef}
\end{eqnarray}
Furthermore, the matter source stress-energy tensor is written with respect to this perturbation in
the form:
\begin{equation}
T_{\mu\nu}^{mat}=\uz T_{\mu\nu}^{mat}+ \uu T_{\mu\nu}^{mat}
\label{eq:tmunupert1}
\end{equation}
As a consequence,  the equation
\begin{equation}
L'(R) R_{(\mu\nu)}(\Gamma)-\frac{1}{2}L(R)g_{\mu \nu }=\kappa
T_{\mu \nu }^{mat}  \label{eq:ffv11}
\end{equation}
 can be written under the
form\footnote{We have taken into account the fact that, on the background,
\begin{equation}
L'(\uz R) \uz R_{(\mu\nu)}(\uz \Gamma)-\frac{1}{2}L(\uz R)\uz
g_{\mu \nu }=\kappa \uz T_{\mu \nu }^{mat}
\label{eq:fieldbackground}
\end{equation}
}
\begin{equation}
L'(\uz R) \uu R_{\mu\nu}+L'(\uu R) \uz R_{\mu\nu}-\frac{1}{2} \uu
g_{\mu\nu}L(\uz R)-\frac{1}{2}  \uz g_{\mu\nu}L(\uu R)=\kappa \uu
T_{\mu\nu}^{mat} \label{eq:fieldp1}
\end{equation}
The  Ricci curvature $\uz R_{\mu\nu}$ (and the corresponding Ricci
scalar $\uz R$) refer to the background solution; in terms of the
perturbation of this solution we may write
\begin{equation}
R_{\mu\nu} = \uz R_{\mu\nu} + \uu R_{\mu\nu} \label{eq:riccip1}
\end{equation}
and
\begin{equation}
R= \uz R + \uu R +  \uu g^{\mu\nu} \uz R_{\mu\nu}
\label{eq:riccisp1}
\end{equation}
So, in order to explicitly write the perturbed field equations
(\ref{eq:fieldp1}) we have to evaluate the perturbed Ricci curvature
and scalar in terms of the fields $g$ and $\Gamma$. In general, we have \cite{Barraco}:
\begin{equation}
R_{\mu\nu}(\Gamma)-R_{\mu\nu}(g) =
\nabla_{(\mu}Q_{\nu)\alpha}^{\alpha}-\nabla_{\alpha}Q^{\alpha}_{\mu\nu}+Q_{\beta(\mu}^{\alpha}Q_{\nu)\alpha}^{\beta}
-Q_{\mu\nu}^{\alpha}Q_{\alpha\beta}^{\beta} \label{eq:riccigamma1}
\end{equation}
where
\begin{equation}
Q_{\mu\nu}^{\alpha} \doteq \left\{^\alpha _{\mu\nu}
\right\}-\Gamma_{\mu\nu}^{\alpha} = \frac{1}{2}g^{\alpha\beta}
\left(\nabla_\mu g_{\nu\beta}+\nabla_\nu g_{\mu\beta}-\nabla_\beta
g_{\mu\nu} \right) \label{eq:defq1}
\end{equation}
in terms of the Christoffel symbols
\begin{equation}
 \left\{^\alpha _{\mu\nu}
\right\} = \frac{1}{2}g^{\alpha\beta} \left( g_{\nu\beta,\mu}+
g_{\mu\beta,\nu}- g_{\mu\nu,\beta} \right) \label{eq:defq111}
\end{equation}
Notice that $\nabla_\mu \doteq \nabla_\mu^\Gamma$ here and
henceforth and we denote moreover with $g_{\mu\beta,\nu}$ the partial
derivative $\partial_\nu g_{\mu\beta}$. The second set of field equations
\begin{equation}
\nabla _{\alpha }(\sqrt{g} \; [ L^{\prime }(R)] \; g^{\mu \nu })=0
\label{ffv21}
\end{equation}
can be now written in the form
\begin{equation}
\nabla_\alpha g_{\mu\nu}=b_\alpha g_{\mu\nu} \label{eq:ffv22}
\end{equation}
We have here defined:
\begin{equation}
b_\alpha \doteq -\nabla_\alpha \left[\ln L'(R) \right]
\label{eq:defbalpha}
\end{equation}
>From  the structural equation (\ref{struct}) and from (\ref{eq:defbalpha}), we obtain then
\begin{equation}
b_\alpha \doteq -\kappa
\frac{L''(R)}{L'(R)}\frac{ \tau_{,\alpha}}{L''(R) R -L'(R)}
\label{eq:defalpha1}
\end{equation}
As a consequence, from equation (\ref{eq:defq1}) we may write
\begin{equation}
Q_{\mu\nu}^{\alpha} = \frac{1}{2}g^{\alpha\beta} \left(b_\mu
g_{\nu\beta}+b_\nu g_{\mu\beta}-b_\beta g_{\mu\nu} \right)
\label{eq:defq2}
\end{equation}
The expression of the Ricci tensor of the affine
connection reads then
\begin{equation}
R_{\mu\nu}(\Gamma)=R_{\mu\nu}(g)+\nabla_{(\mu}b_{\nu)}-\frac{1}{2}b_\mu
b_\nu+g_{\mu\nu}b_\alpha b^\alpha+\frac{1}{2}\nabla_\alpha
b^\alpha g_{\mu \nu}=R_{\mu\nu}(g)+B_{\mu\nu}  \label{eq:riccigamma2}
\end{equation}
by introducing the tensor
\begin{equation}
B_{\mu\nu}\doteq \nabla_{(\mu}b_{\nu)}-\frac{1}{2}b_\mu
b_\nu+g_{\mu\nu}b_\alpha b^\alpha+\frac{1}{2}\nabla_\alpha
b^\alpha  g_{\mu \nu} \label{eq:defBmunu1}
\end{equation}
This expression (\ref{eq:riccigamma2}) holds in the exact theory, so the task of writing its
linear approximation is fulfilled by separately approximating the
metric Ricci tensor $R_{\mu\nu}(g)$ and the $B_{\mu\nu}$ tensor;
the latter, in particular, depends on the analytic expression of
$L(R)$.\\
The perturbation of the metric Ricci tensor is given by (see
\cite{Barraco, straumann04}):
\begin{equation}
\uu R_{\mu\nu}(g)=\frac{1}{2} \uz g^{\alpha\beta} \left(\uu
g_{\beta\mu|\nu\alpha}-\uu g_{\alpha\beta|\mu\nu}+\uu
g_{\beta\nu|\mu\alpha}-\uu g_{\mu\nu|\beta\alpha} \right)
\label{eq:riccilin1}
\end{equation}
where $_|$ stands for the (metric) covariant derivative with
respect to the background\footnote{The covariant derivative
defined by$_|$ is such that $\uz g_{\mu\nu|\alpha}=0$.}.
By perturbing the $B_{\mu\nu}$ tensor, we obtain
\begin{equation}
B_{\mu\nu}=\uz B_{\mu\nu}+ \uu B_{\mu\nu} \label{eq:bmunu1}
\end{equation}
where
\begin{equation}
\uz B_{\mu\nu}=\uz b_{(\mu;\nu)}-\frac{1}{2}\uz b_\mu \uz
b_\nu+\uz g_{\mu\nu}\uz b_\alpha \uz b^\alpha+\frac{1}{2} \uz
b^\alpha_{;\alpha} g_{\mu \nu} \label{eq:bmunu2}
\end{equation}
and
\begin{eqnarray}
\uu B_{\mu\nu} & = & \uu b_{\mu,\nu}- \uz \Gamma_{\ \nu\mu}^\alpha
\uu b_\alpha - \uu \Gamma_{\ \nu\mu}^\alpha \uz
b_\alpha-\frac{1}{2}\uz b_\mu \uu b_\nu-\frac{1}{2}\uu b_\mu \uz
b_\nu+ \nonumber \\ & & + h_{\mu\nu} \uz b_\alpha \uz b^\alpha+\uz
g_{\mu\nu} \uz b_\alpha \uu b^\alpha +\uz g_{\mu\nu} \uu b_\alpha
\uz b^\alpha+\frac{1}{2}\uz g_{\mu\nu} \uu b^\alpha_{,\alpha}+
\nonumber \\ & & + \frac{1}{2}\uz g_{\mu\nu}\uz \Gamma_{\
\alpha\gamma}^\alpha \uu b^\gamma + \frac{1}{2}\uz g_{\mu\nu}\uu
\Gamma_{\ \alpha\gamma}^\alpha \uz
b^\gamma+\frac{1}{2}h_{\mu\nu}\uz b^\alpha_{;\alpha}
\label{eq:bmunu3}
\end{eqnarray}
Notice that $_{;}$ stands here for the covariant derivative with respect
to the unperturbed connection $\uz \Gamma$.


\subsection{Perturbation of flat space-time}\label{ssec:perflat}

We recall here that field equations are
\begin{eqnarray}
L'(R) R_{(\mu\nu)}(\Gamma) &-& \frac{1}{2}L(R)g_{\mu \nu }  =
\kappa T_{\mu \nu }^{mat}  \label{eq:flateq1} \\
\nabla_\alpha g_{\mu\nu} & = & b_\alpha g_{\mu\nu}
\label{eq:flateq2}
\end{eqnarray}
where $b_\alpha$ is defined by formula (\ref{eq:defalpha1}). It is easy to check that the pair $g_{\mu\nu}=\eta_{\mu\nu}$,
$\Gamma=0$, i.e. the Minkowski flat space-time is a solution of
the field equations (\ref{eq:flateq1}), (\ref{eq:flateq2}) iff
\begin{equation}
L(R=0)=0 \label{eq:lrzero}
\end{equation}
 In fact, in vacuum $T \equiv 0$, hence $b_\alpha
=0$\footnote{Notice that in order to have a well posed definition
of $b_\alpha$, we must have $L'(R) \neq 0$, and $L''(R) R -L'(R)
\neq 0$.}.\\
Now, we have to solve the field equations in terms of a
perturbation of the Minkowski flat solution. In particular, we look for solutions in the form
\begin{eqnarray}
\Gamma_{\ \mu\nu}^{\alpha} & = & ^{(0)}\Gamma_{\
\mu\nu}^{\alpha}+^{(1)}\Gamma_{\ \mu\nu}^{\alpha}=^{(1)}\Gamma_{\
\mu\nu}^{\alpha}
\label{eq:eqflat4} \\
g_{\mu\nu} & = &
^{(0)}g_{\mu\nu}+^{(1)}g_{\mu\nu}=\eta_{\mu\nu}+^{(1)}g_{\mu\nu}
\label{eq:eqflat5}
\end{eqnarray}
In what follows we use Cartesian coordinates adapted to the
background metric $^{(0)}g_{\mu\nu}=\eta_{\mu\nu}$; furthermore,
the latter is used to raise and lower indices. The matter source stress-energy tensor is written in the form:
\begin{equation}
T_{\mu\nu}^{mat}=\uz T_{\mu\nu}^{mat}+ \uu T_{\mu\nu}^{mat}=\uu
T_{\mu\nu}^{mat} \label{eq:eqflat6}
\end{equation}
The  Ricci curvature is written in the form
\begin{equation}
R_{\mu\nu} = \uz R_{\mu\nu} + \uu R_{\mu\nu} =  \uu R_{\mu\nu}
\label{eq:eqflat7}
\end{equation}
and the corresponding Ricci Scalar (owing to $\uz R_{\mu \nu} (\eta)=0$):
\begin{equation}
R= \uz R + \uu R +  \uu g^{\mu\nu} \uz R_{\mu\nu}=\uu R
\label{eq:eqflat8}
\end{equation}
Notice that both the Ricci curvature and the Ricci scalar, when it
is not explicitly stated (like in the above equations), refer to
the connection $\Gamma$. As a consequence,  equation (\ref{eq:fieldbackground}) can be
written in the form
\begin{equation}
L'(0) \uu R_{\mu\nu}(\Gamma)-\frac{1}{2}  \eta_{\mu \nu} L(\uu
R)=\kappa \uu T_{\mu\nu}^{mat} \label{eq:eqflat9}
\end{equation}
>From eq. (\ref{eq:riccigamma2}), the perturbed Ricci tensor is
made of two contributions:
\begin{equation}
\uu R_{\mu\nu}(\Gamma)= \uu R_{\mu\nu}(g)+\uu B_{\mu\nu}
\label{eq:eqflat10}
\end{equation}
The perturbation of the metric part of the Ricci tensor is
obtained by replacing the covariant derivative $_|$ with the
ordinary derivative in (\ref{eq:riccilin1}), since our background
is Minkwowski flat space-time:
\begin{equation}
\uu R_{\mu\nu}(g)=\frac{1}{2} \uz g^{\alpha\beta} \left(\uu
g_{\beta\mu,\nu\alpha}-\uu g_{\alpha\beta,\mu\nu}+\uu
g_{\beta\nu,\mu\alpha}-\uu g_{\mu\nu,\beta\alpha} \right)
\label{eq:eqflat11}
\end{equation}
On the other hand, since on the background one has $\uz b_\alpha \equiv
0$, the perturbation of the $B_{\mu\nu}$ tensor reads now as:
\begin{equation}
\uu B_{\mu\nu} = \uu b_{(\mu,\nu)}+\frac{1}{2} \eta_{\mu\nu} \uu
b^\alpha_{,\alpha} \label{eq:eqflat12}
\end{equation}
Hence, the perturbed Ricci tensor turns out to be
\begin{equation}
\uu R_{\mu\nu}(\Gamma)=\frac{1}{2} \eta^{\alpha\beta} \left(\uu
g_{\beta\mu,\nu\alpha}-\uu g_{\alpha\beta,\mu\nu}+\uu
g_{\beta\nu,\mu\alpha}-\uu g_{\mu\nu,\beta\alpha} \right)+\uu
b_{(\mu,\nu)}+\frac{1}{2} \eta_{\mu\nu} \uu b^\alpha_{,\alpha}
\label{eq:eqflat13}
\end{equation}
By exploiting gauge freedom, we may arbitrarily impose the following gauge condition
\begin{equation}
g^{\mu\nu} \Gamma^{\alpha}_{\ \mu\nu}=0 \label{eq:eqflat14}
\end{equation}
which, in linear approximation and by forgetting vanishing terms, becomes:
\begin{equation}
\eta^{\mu\nu} \uu \Gamma^{\alpha}_{\ \mu\nu}=0
\label{eq:eqflat15}
\end{equation}
By taking the linear approximations of eqs.
(\ref{eq:defq1}), (\ref{eq:defq111}), (\ref{eq:defq2}), condition
(\ref{eq:eqflat14}) simply becomes
\begin{equation}
\uu g_{\mu\alpha,}^{\ \ \ \alpha}-\frac{1}{2}\uu g^{\alpha}_{\
\alpha, \mu}+\uu b_\mu=0 \label{eq:eqflat16}
\end{equation}
The gauge condition (\ref{eq:eqflat16}) allows us to write the
perturbed Ricci tensor and the corresponding scalar curvature under
the form
\begin{equation}
\uu R_{\mu\nu}(\Gamma)=-\frac{1}{2} \uu g_{\mu\nu,\alpha}^{\ \ \ \
\alpha}+\frac{1}{2} \eta_{\mu\nu} \uu b^{\alpha}_{\   , \alpha}
\label{eq:eqflat18}
\end{equation}
\begin{equation}
\uu R(\Gamma)=\eta^{\mu\nu} \uu
R_{\mu\nu}(\Gamma)=-\frac{1}{2}\uu g_{\ \mu,\alpha}^{\mu \ \ \
\alpha}+2 \uu b^{\alpha}_{\ , \alpha} \label{eq:eqflat19}
\end{equation}
Now we are in position  to explicitly write the field equation
(\ref{eq:eqflat9}). By taking into account (\ref{eq:lrzero}) ,we may now suppose
that the  function $L(R)$ has the explicit form
\begin{equation}
L(R)=R+\alpha f(R) \label{eq:eqflat20}
\end{equation}
where $\alpha$ is a constant parameter, and $f(R)$ is some
function that  for simplicity we may think to be as a polynomial of degree higher than
one.  A similar analysis holds for thr non-polynomial but still real analytic functions $f(R)$. Consequently, up to linear order we have
\begin{equation}
L'(0)=1 \ \ \ \ L(\uu R) \simeq \uu R \label{eq:eqflat21}
\end{equation}
and field equations become:
\begin{equation}
 \uu R_{\mu\nu}-\frac{1}{2} \eta_{\mu\nu} \uu
R=\kappa \uu T_{\mu\nu}^{mat} \label{eq:eqflat22}
\end{equation}
By substituting the expression of the Ricci tensor and the scalar
curvature (\ref{eq:eqflat18}), (\ref{eq:eqflat19}), we obtain
\begin{equation}
-\frac{1}{2} \uu g_{\mu\nu,\alpha}^{\ \ \ \ \alpha}+\frac{1}{4} \eta_{\mu\nu} \uu g_{\ \mu,\alpha}^{\mu \ \ \
\alpha}=\kappa \uu
T_{\mu\nu}^{mat}+\frac{1}{2} \eta_{\mu\nu} \uu b^{\alpha}_{\ ,
\alpha} \label{eq:eqflat23}
\end{equation}
Now, from (\ref{eq:defalpha1}), up to first order we may write
\begin{equation}
\uu b^{\alpha}_{\ , \alpha} \simeq -\kappa \frac{L''(0)}{\left(
L'(0) \right)^2} \uu T^{mat\ \ \alpha}_{ \ \ \ \ , \ \alpha}
\label{eq:eqflat24}
\end{equation}
Furthermore we may introduce the tensor $\overline{h}_{\mu\nu}$
defined by
\begin{equation}
\overline{h}_{\mu\nu} \doteq \uu g_{\mu\nu}-\frac{1}{2} \eta_{\mu\nu} \uu g^\alpha_{\ \alpha} \label{eq:eqflat25}
\end{equation}
Then by means of (\ref{eq:eqflat24}) and (\ref{eq:eqflat25}) the
field equations (\ref{eq:eqflat23}) simplify to
\begin{equation}
\square \left[\overline{h}_{\mu\nu} - \eta_{\mu\nu} \kappa L''(0)
\uu \tau \right]=-2\kappa \uu T_{\mu\nu}^{mat}
\label{eq:eqflat26}
\end{equation}
If we set
\begin{equation}
L(R) = R + \alpha f(R)=R+\alpha R^2+P(R) \label{eq:eqflat27}
\end{equation}
where $P(R)$ is a polynomial of degree higher than 2, the field
equations (\ref{eq:eqflat26}) can consequently be written in the form
\begin{equation}
\square \left[\overline{h}_{\mu\nu}-2\eta_{\mu\nu} \kappa \alpha
\uu T \right]=-2\kappa \uu T_{\mu\nu}^{mat}
\label{eq:flatfieldeq2}
\end{equation}
By setting
\begin{equation}
\mathcal{H}_{\mu\nu} \doteq \overline{h}_{\mu\nu}-2\eta_{\mu\nu}
\kappa \alpha \tau^{mat} \label{eq:eqflat28}
\end{equation}
the field equations take the simple expression
\begin{equation}
\square \mathcal{H}_{\mu\nu}=-2\kappa \uu T_{\mu\nu}^{mat}
\label{eq:eqflat29}
\end{equation}
The solution of (\ref{eq:eqflat29}) can now be written  in terms of
retarded potentials:
\begin{equation}
\mathcal{H}_{\mu\nu}=4\frac{G}{c^{4}}\int \frac{T^{mat}_{\mu \nu }(t-|\mathbf{x
}-\mathbf{x^{\prime }}|/c,\mathbf{x^{\prime }})}{|\mathbf{x}-\mathbf{
x^{\prime }}|}d^{3}x^{\prime }  \label{eq:eqflat30}
\end{equation}
where we have explicitly written $\kappa=8 \pi  G /c^4$. Hence
\begin{equation}
\overline{h}_{\mu\nu}=4\frac{G}{c^{4}}\int \frac{T^{mat}_{\mu \nu }(t-|\mathbf{x
}-\mathbf{x^{\prime }}|/c,\mathbf{x^{\prime }})}{|\mathbf{x}-\mathbf{
x^{\prime }}|}d^{3}x^{\prime }+\frac{16 \pi G}{c^4}\eta_{\mu\nu} \
\alpha \tau^{mat} \label{eq:eqflat31}
\end{equation}
>From the above calculations, which have been performed step by step to  exactly clarify what happens, it turns out that
the first order perturbation does not influence the form of field equations (\ref{eq:eqflat9}), while it enters into the definition of
the perturbed Ricci tensor  (\ref{eq:eqflat10}) and consequently of the  scalar curvature. \\
Specifying to the case of the Lagrangian (\ref{eq:eqflat27}) we see that the solution of the perturbed field equations, written in terms of the retarded potentials, contains two terms: i) the first one, in the weak field approximation, reduces to the standard Newtonian potential,  ii) the second one is related to the Lagrangian chosen for the alternative theory of gravity. In particular it vanishes in the limit $\alpha \rightarrow 0$, i.e. exactly reproducing the weak field limit of standard General Relativity. It is thus clear that $\alpha$ can be identified with a scale parameter, vanishing at small (solar system) scales and consequently reproducing General Relativity.  The same reasoning can be done by supposing that all the term $\alpha f(R)$ becomes in fact irrelevant at solar system scales. \\


\subsection{Perturbation of the de Sitter space-time} \label{ssec:perdesitter}

The calculation performed in the previous sub-section can be generalized to the case of space-times which do not admit a Minkowski  background solution. However, as we have seen in the previous section, having a Minkowski solution heavily constrains the available Lagrangians, since it  implies that $L(R=0)=0$: in particular, these Lagrangians are not  interesting for cosmological applications (see  \cite{carrol2} and  \cite{Barraco2}). This case was already studied in \cite{Barraco2}, where it is shown that theories with singular $L(R)$ and $\frac{d^2 L}{dR^2} (^{(0)}R)=0$ provide the correct Newtonian limit and they are good candidates to explain the cosmic acceleration. \\
We want hereafter to comment this case and to apply it to the particular Lagrangian $L(R)=R + \alpha f(R)$. We skip calculations as they can be reproduced step by step following the headlines of the previous chapter and moreover they have been already performed in \cite{Barraco2}. We consider as a background metric the (anti) de Sitter metric:
\begin{equation}
\uz g=- dt^2+e^{2 t \sqrt{\Lambda \over 3}}(dr^2+r^2d\vartheta^2+r^2\sin^2
\vartheta d\varphi^2) \label{eq:metricasitter}
\end{equation}
which satisfies the field equations:
\begin{equation}
\uz R_{\mu \nu}=- \Lambda {\uz g}_{\mu \nu}
\end{equation}
Considering the Lagrangian $L(R)=R + \alpha f(R)$ we obtain that the resulting Newtonian potential is (see \cite{Barraco2} for details on calculations):
\begin{equation}
V(\mathbf{x})= e^{2 t \sqrt{\Lambda \over 3}} C \int \frac{\rho(\mathbf{x}^\prime) \exp(- |\mathbf{x}-\mathbf{
x^{\prime }}|  e^{2 t \sqrt{\Lambda \over 3}})}{\mathbf{x}-\mathbf{
x^{\prime }}}   d^{3}x^{\prime }+A \rho (\mathbf{
x^{\prime }}) \label{eq:eqsitter31}
\end{equation}
where $\rho( {\mathbf{x}})$ represents the energy density while:
\[
\cases{
A=\alpha \frac{\kappa {\uz f}^{\prime \prime}}{2 {\uz L}^\prime(4 \Lambda \alpha {\uz f}^{\prime \prime}+ {\uz L}^{ \prime})} \cr
\cr
C=\frac{8 \Lambda \alpha {\uz f}^{\prime \prime}  {\uz L}^{\prime}+ ({\uz L}^{ \prime})^2}
{ ({\uz L}^{ \prime})^2 (4 \Lambda \alpha {\uz f}^{\prime \prime}+ {\uz L}^{ \prime})}
}
\]
Also in this case it is evident that in the limit $\alpha \rightarrow 0$, for any current experiment and observation, the first term in (\ref{eq:eqsitter31}) reduces to  the standard Newtonian  potential \cite{Barraco2}, and once again  we obtain the weak field limit of standard General Relativity.  Moreover $\alpha$ naturally behaves as
a Post-Newtonian parameter in the potential, which is supposed to vanish at small scales. Once more $\alpha$ behaves like a scale parameter and the accordance with experimental results is supported.

\section{Conclusions}

In this paper we have  shown that, both in the case of vacuum
universes and in the case of matter universes, solar system
experiments can be theoretically explained and reproduced in the
framework of alternative theories of Gravity for specific classes of Lagrangians. The gravitational
potential of alternative theories of Gravity reduces, under
suitable hypotheses, to the standard Newtonian potential at the solar
system scale. This has been proven both in the case of vacuum and
matter universes (with a flat or an (anti) de-Sitter background).
Gravitational effects due to the (alternative) form of the
Lagrangian generate Post-Newtonian parameters appearing in the
gravitational potential, which vanish when the corrections to the
standard Hilbert Lagrangian are cancelled. Moreover we stress
that these corrections are negligible when we consider values of
the scale parameter $\alpha$ which is necessary to explain cosmic
acceleration (see e.g. \cite{carrol2}). These contributions become
however relevant when considering larger scales (cosmology
\cite{palatinifR} and, hopefully, galactic scales). This implies
that higher order corrections to the standard Hilbert-Einstein
theory could behave as a scale effect, ruled by a scale parameter
which vanishes at solar system scales.
General Relativity is consequently reproduced at the solar system scale, as it has to be
surely expected.     \\
The results obtained  here slightly differ from some results already
presented in literature \cite{palatinifR} and in particular from
recent results obtained  by G.J. Olmo, (see
again \cite {palatinifR}). It was there argued that only small corrections to the
Hilbert-Einstein Lagrangian can pass the solar system experiments.
However, calculations were there performed by means of a conformal
transformation on a flat  Minkowski background spacetime. We have
here shown that in the particular case of a flat spacetime, the
theory $1 \over R$ is not viable, owing to the conditions (\ref
{eq:lrzero}). This implies that the results obtained by G.J. Olmo
does not exclude the reliability of  $1 \over R$-like theories,
which should however  be examined in the (anti)de-Sitter background
framework. In fact $R=0$ is singular for all Lagrangians which contain inverse
powers or logarithms. Moreover,  it is not at all evident
that our universe should be asymptotically flat. We have here proven the accordance of such theories
with solar system experiments, at least when the scale
(post-Newtonian) parameter becomes small enough.


\section{Acknowledgements}
\noindent  We are very grateful to Prof. S.D. Odintsov and Dr. Andrzej Borowiec for helpful comments and suggestions.



\begin{thebibliography}{99}

\bibitem{Perlmu} Perlmutter, S. et al. 1999, ApJ, 517, 565 (astro-ph/9812133 ); Perlmutter el al. Nature 404 (2000) 955 Astroph. J. 517 (1999) 565, Riess, A. G. et al. 1998, AJ, 116, 1009 (astro-ph/9805201), Spergel, D. N. et al. 2003, ApJS, 148, 175, Verde, L. et al. 2002, MNRAS, 335, 432.

\bibitem{dvali} Dvali, G., Gabadadze, G.,  Porrati, M. 2000, Phys. Lett. B485, 208 (hep-th/0005016); Dvali, G., Gruzinov, A.,  Zaldarriaga, M. 2003, Phys. Rev. D, 68, 24012;   Freese, K.  Lewis, M. 2002, Phys. Lett. B540, 1 (astro-ph/0201229);  I.Ya.Aref'eva, L.V.Joukovskaya (hep-th/0504200); I.Ya.Aref'eva, A.S.Koshelev, S.Yu.Vernov (astro-ph/0412619), S. Nojiri and S.D. Odintsov (hep-th/0307071).

\bibitem{carrol2} S.M. Carroll, V. Duvvuri, M. Trodden and M.S. Turner,
(astro-ph/0306438); S. Capozziello, S. Carloni and A. Troisi, "Recent
Research
Developments in Astronomy and Astrophysics" -RSP/AA/21-2003
(astro-ph/0303041);
S. Capozziello, Int. J. Mod. Phys. D11, 483 (2002); S. Capozziello, V.F.
Cardone, S. Carloni and A. Troisi, Int. J. Mod. Phys. D12, 1969 (2003).

\bibitem{staro} A. A. Starobinsky, Phys, Lett. B 91 ($1980$) 99.

\bibitem{metricfR}
S. Nojiri and S.D. Odintsov, Phys. Rev. D68, 123512 (2003)
(hep-th/0307288);  Gen. Rel. Grav. 36 (2004) 1765 (hep-th/0308176); Mod.Phys.Lett.A 19 (2004)627
(hep-th/0310045), G. J. Olmo, (gr-qc/0505135).


\bibitem{palatinifR} D.N. Vollick, Phys. Rev. D68, 063510 (2003); G. Allemandi, A. Borowiec and M. Francaviglia,  Phys.Rev. D
{\bf 70}, 043524 (2004), (hep-th/0403264); G.J. Olmo  gr-qc/0505101; G.J. Olmo and W. Komp, (gr-qc/0403092); X.H. Meng and P. Wang, Class. Quant. Grav. 20, 4949-4962 (2003) (astro-ph/0307354); X.H. Meng and P. Wang, Class. Quant. Grav. 21, 951-960 (2004) (astro-ph/0308031); G.M. Kremer and D.S.M. Alves, (gr-qc/0404082), G. J. Olmo, (gr-qc/0505136), G. Allemandi, A. Borowiec, M. Francaviglia and S.D. Odintsov, (gr-qc/0504057);

\bibitem{zerbo} G. Cognola, E. Elizalde, S. Nojiri, S.D. Odintsov and S. Zerbini, JCAP0502, 010, 2005 (hep-th/0501096);

\bibitem{solar} C. Will, Living Reviews in Relativity, lrr-2001-4 (www.livingreviews.org);


\bibitem{solarmet}  T.Chiba, Phys.Lett. B575 (2003) 1-3 (astro-ph/0307338); A. D. Dolgov and M. Kawasaki, Phys.Lett. B573 (2003) 1-4 (astro-ph/0307285); S.Nojiri and S.D.Odintsov, Phys.Rev. D68 (2003) 123512 (hep-th/0307288), S. Capozziello (gr-qc/0412088);

\bibitem{solarpal} D. Barraco, V. H. Hamity, Gen. Rel. Grav. Vol 25, 461-471; D. Barraco, V. H. Hamity, Gen. Rel. Grav.ÊÊ Vol. 28, 339-345 (1996); X.H.Meng and P.Wang, Class.Quant.Grav. 20 (2003) 4949-4962

(astro-ph/0307354); Class.Quant.Grav. 21 (2004) 951-960 (astro-ph/0308031);  (astro-ph/0308284);  D. N. Vollick,  (gr-qc/0312041).

\bibitem{Barraco}  D. Barraco, V. H. Hamity, H. Vucetich, Gen. Rel. Grav. 34 (4) (2002), 533-547.

\bibitem{Barraco2}  A. E. Dominguez, D. E. Barraco; Phys.Rev. D70 (2004) 043505 (gr-qc/0408069).

\bibitem{capogal} S. Capozziello, V.F. Cardone, S. Carloni, A. Troisi;  Phys.Lett. A326 (2004) 292-296, .

\bibitem{MGnew} G. Allemandi, M.L. Ruggiero, {\it in preparation}

\bibitem{buchdahl} H.A. Buchdahl, J. Phys. A {\bf 12} (8) (1979), 1229; \\ H.A. Buchdahl, Proc. Camb. Phil. Soc. {\bf 68} (1960), 396.

\bibitem{FFV} M.Ferraris, M.Francaviglia and I.Volovich, Nouvo Cim. B108 (1993) 1313 (gr-qc/9303007);
M.Ferraris, M.Francaviglia and I.Volovich, Class. Quant.Grav. 11 (1994) 1505.

\bibitem{he} S.W. Hawking, G.F.R. Ellis, \textit{The large scale structure of
space-time}, Cambridge University Press, Cambridge (1973)

\bibitem{ruffi}  C. Ohanian Hans,  R. Ruffini, \textit{Gravitation and
Spacetime}, W.W. Norton and Company, New York (1994)

\bibitem{straumann04} N. Straumann, \textit{General Relativity},
Springer, Berlin (2004)

\end{thebibliography}
\end{document}